\documentclass[twocolumn,prc,showpacs]{revtex4}
\usepackage{graphicx}
\begin{document}

\title{Particle decay branching ratios for states of astrophysical importance in $^{19}$Ne}

\author{D. W. Visser}
\altaffiliation[Present address: ]{Physics Division, Oak Ridge
National Laboratory, Oak Ridge, Tennessee 37831}
\email{dale.visser@mailaps.org}
\author{J. A. Caggiano}
\author{R. Lewis}
\author{W. B. Handler}
\altaffiliation[Present address: ]{Physics and Astronomy
Department, University of Western Ontario, London, Ontario, Canada
N6A 3K7}
\author{A. Parikh}
\author{P. D. Parker}
\affiliation{A. W. Wright Nuclear Structure Laboratory, Yale
University, New Haven, Connecticut 06520-8124}

\date{\today}

\begin{abstract}
We have measured proton and alpha-particle branching ratios of
states formed using the $^{19}$F($^{3}$He,$t$)$^{19}$Ne$^*$
reaction at a beam energy of 25 MeV. These ratios have a large
impact on the astrophysical reaction rates of
$^{15}$O($\alpha$,$\gamma$)$^{19}$Ne,
$^{18}$F(p,$\gamma$)$^{19}$Ne and $^{18}$F(p,$\alpha$)$^{15}$O,
which are of interest in understanding energy generation in x-ray
bursts and in interpreting anticipated $\gamma$-ray observations
of novae. We detected decay protons and alpha particles using a
silicon detector array in coincidence with tritons measured in the
focal plane detector of our Enge split-pole spectrograph. The
silicon array consists of five strip detectors of the type used in
the Louvain-Edinburgh Detector Array, subtending angles from
$130^{\circ}$ to $165^{\circ}$ with approximately 14$\%$ lab
efficiency. The correlation angular distributions give additional
confidence in some prior spin-parity assignments that were based
on gamma branchings. We measure $\Gamma_p/\Gamma=0.387\pm0.016$
for the 665 keV proton resonance, which agrees well with the
direct measurement of Bardayan \emph{et al.}.
\end{abstract}


\maketitle

\section{Introduction}
States in $^{19}$Ne corresponding to $^{15}$O+$\alpha$
(Q$_{\alpha\gamma}$=3529.4 keV) and $^{18}$F+p
(Q$_{p\gamma}$=6411.2 keV) resonances play an important role in
explosive hydrogen burning. In energetic events known as x-ray
bursts, $^{15}$O($\alpha$,$\gamma$)$^{19}$Ne has long been thought
to be a possible pathway from the hotCNO energy-generation
bottleneck into the more exoergic hydrogen-burning processes above
A=19, such as the NeNa cycle and the rp process \cite{wiescher}.
In novae, the $\beta$$^{+}$ decay of $^{18}$F is expected to be
the prime source of the 511 keV $\gamma$-ray line, and of the
sub-511-keV continuum $\gamma$-ray flux, observable by orbiting
$\gamma$-ray observatories such as INTEGRAL \cite{hernanz}. The
final yield of $^{18}$F and its associated radiation is sensitive
to the reaction rates of $^{18}$F(p,$\gamma$)$^{19}$Ne and
$^{18}$F(p,$\alpha$)$^{15}$O \cite{coc,iliadis}.

The level density at relevant energies above the proton and alpha
thresholds is low, so that the nuclear reactions mostly take place
through isolated and narrow resonances. The temperature-dependent
resonant reaction rates may be calculated by summing over
resonances \cite{cauldrons}:
\begin{equation}\label{reactrate}
  <\sigma v>=\sum_R\left(\frac{2\pi}{\mu
  kT}\right)^{3/2}\hbar^2\left(\omega\gamma\right)_R
  exp\left(-\frac{E_R}{kT}\right),
\end{equation}
where ${\mu}$ is the reduced mass, ${k}$ is Boltzmann's constant,
${T}$ is the temperature, ${E_R}$ is the resonance energy and
${\omega\gamma}$ is its strength. The first factor in the
resonance strength, ${\omega}$, is a statistical factor that
depends on the spins in the incoming and outgoing channel, while
${\gamma}$ depends on the width and branching ratios for the
resonance:
\begin{equation}\label{gammafac}
    \gamma=\frac{\Gamma_a}{\Gamma} \frac{\Gamma_b}{\Gamma} \Gamma,
\end{equation}
where ${a}$ and ${b}$ denote the incoming and outgoing particles
(${p}$, ${\alpha}$ or ${\gamma}$) for the reaction $I(a,b)F$.
Thus, the reaction rate has a linear dependence on the branching
ratios.

\section{Experiment}

\begin{figure*}
\includegraphics{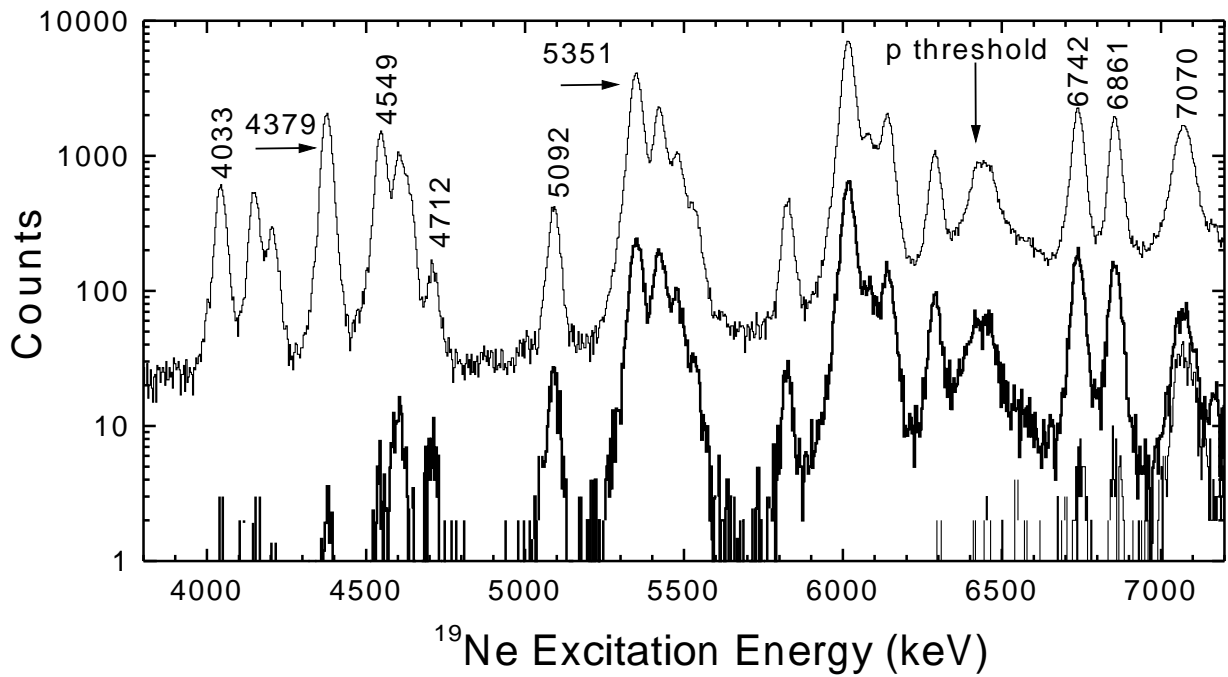}
\caption{\label{spec}The top curve is the triton spectrum from
data set D4, with the abscissa transformed to show $^{19}$Ne
excitation energy. The darker middle curve shows the tritons
coincident with alpha-particles in YLSA, while the lower curve
shows candidate events for tritons coincident with protons in
YLSA, i.e., random contributions are not subtracted. The
$^{15}$O+$\alpha$ threshold lies at $E_x$=3529.4 keV.}
\end{figure*}

Previous experiments have measured astrophysically-important
branching ratios \cite{magnus,utku,laird,rehm,davids} and
resonance strengths \cite{graulich,rehm96,bardhi,bardlo} for
$^{19}$Ne. The goal of the present experiment was to perform an
independent measurement of these branching ratios, and to achieve
higher sensitivity in that measurement by using large-area silicon
strip detectors \cite{leda}. These detectors provided a large
solid angle and an angular resolution in the lab of approximately
2$^{\circ}$ \cite{dale}.

The $^{19}$F($^{3}$He,$t$)$^{19}$Ne reaction was studied using 25
MeV $^{3}$He beams provided by the WNSL ESTU tandem Van de Graaf
accelerator. The beam, with a typical current of 19 pnA, was
incident on 80 ${\mu}$g/cm$^{2}$ of CaF$_{2}$ deposited on a 10
${\mu}$g/cm$^{2}$ carbon foil. The Enge spectrograph was placed at
$\theta=0^{\circ}$ with an aperture solid angle of 12.8 msr, and
reaction products were momentum analyzed using a magnetic field of
approximately 12.9 kG. The focal plane detector \cite{chen}
measured the position (momentum), horizontal angle, $\Delta E$
(energy loss in isobutane) and $E$ (energy loss in a plastic
scintillator). Particle identification (PID) was accomplished
mainly by a gate on energy vs. momentum. Gates on other 2-d event
plots further refined the PID. Tritons were easily distinguished
from the only other incident particles, deuterons. The horizontal
tracking information was used to correct the dominant
$(x|\Theta^3)$ aberration in the spectrograph's focussing.

Protons and alpha particles from the decay of unbound states were
detected by the Yale Lamp Shade Array (YLSA), an array constructed
from five sectors of silicon strip detectors of the type used in
the Louvain-Edinburgh Detector Array \cite{leda}. The segments
were arranged in an axially-symmetric 5-sided ``lamp shade''
configuration covering angles from $\theta=130^{\circ}$ to
$\theta=165^{\circ}$, with approximately 14\% lab efficiency. The
detectors were approximately 500 $\mu$m thick; a bias voltage of
100 V was applied to them. Since it was essential that the beam go
through the center of the chamber, two collimators were used for
tuning: one at the entrance and one at the exit of the scattering
chamber; both were 2 mm in diameter. YLSA was protected from
particles scattered off the up- and downstream collimators by the
detector mount and an aluminum sheet, respectively
\cite{willnima}. After tuning, these collimators were rotated out
of the beam path.

In order to ensure proper functioning of the silicon detectors in
the environment of electrons ejected from the target by the
relatively intense beam, the sides of the detectors closest to the
strips' pn-junctions, which have exposed SiO$_{2}$ between them,
faced away from the target. Also, the target had strong rare-earth
magnets immediately to either side to deflect electrons, and the
target ladder, aluminum sheet and detector mount were biased with
+500 V to attract electrons away from the detector
\cite{willnima}.

A fast trigger was produced by the scintillator in the
spectrograph's focal plane detector. Uninteresting events were
discarded using rough PID based on the $\Delta E$ and $E$ signals.
The ``fast-clear'' (FCLR) feature of the CAEN V785/V775 ADC's and
TDC's was used for this hardware cut, and did not reliably clear
all modules at high trigger rates. This was undiscovered until
after the measurement was made, resulting in the need for
efficiency corrections. Fast discriminators on each of the 80
($5\times 16$) silicon detector channels produced logic pulses
which were passed through fixed 300-nsec delays to produce time
spectra. Decay protons were distinguished from decay alpha
particles by their differing energies.

Once PID was used to select the ($^{3}$He,t) channel and events
with possible decay energies were extracted from the data, the
momentum spectrum of coincident decays (see Fig.~\ref{spec}) was
determined by requiring the time to lie within the peak in the
time spectrum. A background spectrum was produced using candidate
events with measured times outside the peak. This background
spectrum was used to quantify the rate of random coincidences.

The energy calibration for YLSA was determined using a $^{228}$Th
alpha-particle source. The exact location of the discriminator
thresholds varied by strip and was uncertain. The detectors in
YLSA have an estimated 2-$\mu$m dead layer; in simulations,
deposited energies for alpha particles incident on YLSA ranged
from 300 to 600 keV, depending on lab angle and the angle of
incidence. Because of this, the lowest $^{15}O+\alpha$ resonance
for which a reliable branching ratio could be extracted was at
$E_R$=1020 keV ($E_x$=4549 keV). Likewise, the only $^{18}$F+p
resonance with a reliable proton branching ratio was $E_R$=665 keV
($E_x$=7076 keV). Above $E_x$=7076 keV, the triton energy
calibration was uncertain, and states were not well resolved from
one another.

\section{Simultaneous Fit of the Data Sets}

\begin{table*}
    \caption{Branching ratio results below the proton threshold, with error bars given at a 68.27\% confidence level. Unless otherwise stated, J$^\pi$ assignments are from
\cite{tilley}.}
    \label{results1}
    \begin{ruledtabular}
    \begin{tabular}{rrlrrrrr}
    $E_x$ [keV] & $E_R$ [keV]
     & J$^\pi$ & $\Gamma_\alpha/\Gamma$ \footnote{This work.} & $\Gamma_\alpha/\Gamma$ \cite{magnus} & other $\Gamma_\alpha/\Gamma$ & $\Gamma_\alpha/\Gamma$ \cite{davids} & $\Gamma_\alpha/\Gamma$ (Average)\\
    \hline
    4379 & 850 &  7/2$^+$ & ($>$0.0027)\footnote{Tentative. See text for details.} & 0.044$\pm$0.032 & 0.016$\pm$0.005 \cite{rehm} & $<$0.0039 \footnote{90\% confidence.} \\
    4549 & 1020 & (1/2$^-$) \cite{davidson} & 0.06$\pm$0.04 & 0.07$\pm$0.03 & & 0.16$\pm$0.04& 0.09$\pm$0.04\\
    4600 & 1071 & 3/2$^-$ & 0.208$\pm$0.026 \footnote{Unresolved with E$_x$=4635 keV. See text for details.} & 0.25$\pm$0.04 & 0.32$\pm$0.03 \cite{laird}& 0.32$\pm$0.04 & 0.27$\pm$0.05\\
    4712 & 1183 & (5/2$^-$) & 0.69$^{+0.11}_{-0.14}$ & 0.82$\pm$0.15 & & 0.85$\pm$0.04&0.83$\pm$0.05 \\
    5092 & 1563 & 5/2$^+$ & 0.75$^{+0.06}_{-0.07}$ & 0.90$\pm$0.09 & 0.80$\pm$0.10 \cite{rehm}& 0.90$\pm$0.05&0.84$\pm$0.07 \\
    \end{tabular}
    \end{ruledtabular}
\end{table*}

Since all spherical harmonics with non-zero magnetic substates are
suppressed at $\theta=0^{\circ}$, the tritons were detected at
$\theta=0^{\circ}$ in order to simplify the calculation of the
angular correlations. In order to have sufficient counts for
fitting, the detected decay particles were divided into four
angles (4 strips $\times$ 5 detectors per angle). As in
\cite{magnus}, the angular correlation function used for breakup
to $^{15}$O$_{g.s.}$ was
\begin{equation}\label{corr}
    W\left(\theta\right)=\sum_{M,m}
    \vert\langle\frac{1}{2} m; l
    M-m|JM\rangle Y_{l}^{M-m}\left(\theta,\phi\right)\vert^{2},
\end{equation}
where $M$ are the substates of the $^{19}$Ne state with spin $J$,
$p_M=p_{-M}$ are the $J+\frac{1}{2}$ probabilities of substate
populations, and $m=\pm\frac{1}{2}$ are the possible magnetic
substates of the $^{15}$O \cite{satchler}. For $\vert M\vert
>\frac{3}{2}$, $p_M\equiv 0$ because the tritons were detected at
$\theta=0^{\circ}$.

Four separate sets of spectra were analyzed and fit to produce
angular distributions. Each set corresponded to a unique set of
experimental conditions. Higher ADC thresholds were used in the
first data set (D1). The thresholds were lowered for the second
and third data sets (D2,D3), which have slightly different energy
calibrations from one another. For the fourth data set (D4), the
YLSA electronics channels were given a shorter ADC gate than the
focal plane channels, in an attempt to reduce pileup in the
spectra.

The geometric efficiency of each strip in the array was determined
for each resonance using a Monte Carlo simulation of the formation
and breakup of the resonances. The $E_x$=5351 keV state is
well-populated and has an isotropic c.m. angular distribution for
its decay products. It is known to decay almost entirely by
alpha-particle emission. (Calculations give $\Gamma_\gamma/\Gamma
<10^{-3}$ \cite{langanke}; $\omega\gamma=1.69\pm0.14$ eV and
$\Gamma=1.3\pm0.5$ keV have been measured for the analog in
$^{19}$F which is 499 keV closer to its threshold \cite{wilmes}.)
When measuring the branching ratio for this state, it was found
that the result was consistently less than one. The specific
result was dependent on the details of the electronics gating. The
three different electronics configurations resulted in three
different ``coincidence efficiencies'', ranging from
0.581$\pm$0.018 (D4) in the worst case to 0.867$\pm$0.027 (D1).
For each of the runs, the decay data were divided by this factor.

\begin{table*}
    \caption{Branching ratio measurements above the proton threshold.}
    \label{results2}
    \begin{ruledtabular}
    \begin{tabular}{rrlrrrrrr}
    $E_x$ [keV] & $E_R$ [keV]
     & J$^\pi$ & $\Gamma_\alpha/\Gamma$ \footnote{This work.} & $\Gamma_\alpha/\Gamma$ \cite{utku} & $\Gamma_\alpha/\Gamma$ (Average)& $\Gamma_p/\Gamma$ \footnotemark[1] & $\Gamma_p/\Gamma$ \cite{utku} & $\Gamma_p/\Gamma$ (Average)\\
    \hline
    6742 & 330 & 3/2$^-$ & 0.901$^{+0.074}_{-0.031}$ & 1.04$\pm$0.08 & 0.96$^{+0.07}_{-0.04}$ & ($>$0.003)\footnote{Tentative. See text for details.}\\
    6861 & 450 & 7/2$^-$ & 0.932$^{+0.028}_{-0.031}$ & 0.96$\pm$0.08 & 0.935$^{+0.026}_{-0.029}$ & ($>$0.007) \footnotemark[2] & $<$0.025\\
    7076 & 665 & 3/2$^+$ \cite{rehm96,bardhi,graulich} & 0.613 \footnote{Simultaneous fit assumed $\Gamma=\Gamma_\alpha+\Gamma_p$.}
    & 0.64$\pm$0.06 &  & 0.387$\pm$0.016 & 0.37$\pm$0.04 & 0.385$\pm$0.015 \\
    \end{tabular}
    \end{ruledtabular}
\end{table*}

For E$_x$=4379 keV, the lab energies of the alpha-particles
extended below the thresholds for the detector. The exact energy
of these thresholds varies with detector strip and is not
precisely known. If one (falsely) assumes that all decay alpha
particles were detected for this state, a 90\% confidence interval
for $\Gamma_\alpha/\Gamma$ of (0.0027,0.0143) is found. This is
interpreted as meaning $\Gamma_\alpha/\Gamma>$0.0027 with a
probability greater than 90\%. Lower limits of 0.003 and 0.007 for
$\Gamma_p/\Gamma$ of the E$_x$=6742 and 6861 keV states,
respectively, were determined in the same way. In these latter two
cases, only data set D4 supports a non-zero result, which is cause
for skepticism, especially since calculations based on the
measured mirror state in $^{19}$F \cite{utku} and a direct
measurement of $\Gamma_p$=2.22$\pm$0.69 eV \cite{bardlo} estimate
$\Gamma_p/\Gamma$=8.2$\times10^{-4}$ for the E$_x$=6742 keV state
\cite{utku}. The ratio extracted from D4 is significantly
different only for these two states.

As can be seen in Fig.~\ref{spec}, the energy (E$_x$) resolution
of the triton spectrum was about 40 keV, so the peaks for the
states at E$_x$=4600 and 4635 keV were not well resolved.  Since
the E$_x$=4635 keV state is not expected to emit significant
numbers of alpha-particles (due to its $l=7$ angular momentum
barrier), in our analysis of the coincidence data for these two
states we attributed all of the measured decay alpha-particles to
the E$_x$=4600 keV state.

\begin{figure}
\includegraphics[width=3.25in]{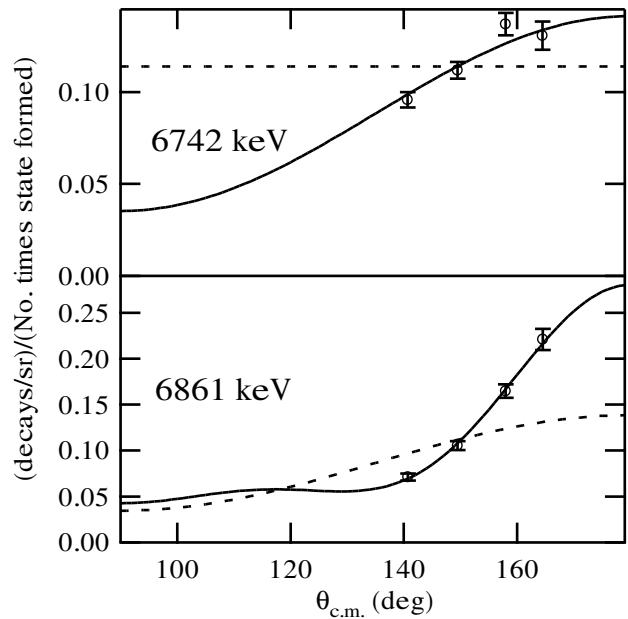}
\caption{\label{angdist}Angular distributions of decay
alpha-particles. (a) $E_x$=6742 keV, solid curve for
$J^\pi$=$\frac{3}{2}^-$, dashed curve for $J^\pi$=$\frac{1}{2}^-$.
(b) $E_x$=6861 keV, solid curve for $J^\pi$=$\frac{7}{2}^-$,
dashed curve for $J^\pi$=$\frac{3}{2}^-$.}
\end{figure}

For the 330-keV proton resonance at $E_x$=6742 keV, we measured
the spin and parity using the angular distribution of its decay
alpha-particles. The $^{20}$Ne($^3$He,$\alpha$)$^{19}$Ne angular
distribution measured for this state shows $l=1$ character,
implying negative parity and $J=\frac{1}{2}$ or $\frac{3}{2}$
\cite{garrett}. $J^{\pi}=\frac{1}{2}^-$ would imply an isotropic
correlation function for the decay alpha-particles, but
Fig.~\ref{angdist}a shows that $J^{\pi}=\frac{3}{2}^-$ results in
a much better fit to the data.

A previous high-resolution measurement has shown that the
($^3$He,t) reaction populates the 450 keV proton resonance
\cite{utku}, and not an $l=1$ resonance (isospin mirror of the
J$^{\pi}$=$\frac{3}{2}^{-}$, E$_x$=6891 keV state in $^{19}$F)
predicted to lie near 430 keV \cite{shu}. The decay alpha-particle
angular distribution in Fig.~\ref{angdist}b shows that a spin
assignment of $\frac{7}{2}^{-}$ is the appropriate choice for the
peak at this location in our spectrum.

The measurement of $\Gamma_p/\Gamma=0.387\pm0.016$ for the 665-keV
proton resonance is in agreement with the previously measured
value \cite{utku}. It disagrees with the branching ratio implied
by the direct measurement performed at Louvain-la-Neuve
\cite{graulich}, but agrees with the measurements made at Oak
Ridge National Laboratory \cite{bardhi}.

\section{Conclusions}

The branching ratio measurements presented here are compared with
previous measurements in Tables \ref{results1} and \ref{results2}.
In cases where other measured values exist, weighted averages are
given. The measurement provides support for the
$J^\pi$=$\frac{3}{2}^-$ assignment for the 330-keV
$^{18}$F(p,$\gamma$) resonance, which dominates the reaction rate
above 450 MK \cite{utku}. The 450-keV $^{18}$F+p resonance is also
confirmed to have $J^\pi$=$\frac{7}{2}^-$, re-affirming its weak
role in astrophysical processes \cite{utku}. The measured
branching ratio for the 665-keV $^{18}$F+p resonance, which
strongly affects both $^{18}$F(p,$\gamma$) and
$^{18}$F(p,$\alpha$) astrophysical rates \cite{utku}, differs from
one recent direct measurement \cite{graulich} and agrees with
another \cite{bardhi}. Therefore, these reaction rates can now be
calculated with greater confidence.

\section{Acknowledgements}

This work was supported under US Department of Energy Grant
No.DE-FG02-91ER-40609.

\bibliography{article}

\end{document}